\documentstyle[12pt,aaspp4,psfig]{article}
\newcommand{\ms}{\mbox{m s$^{-1}~$}}
\newcommand{\kms}{\mbox{km s$^{-1}~$}}
\newcommand{\msun}{M$_{\odot}$}

\newcommand{\mjup}{M$_{\rm JUP}~$}

\newcommand{\msini}{$M \sin i~$}

\lefthead{Marcy{\it et~al.\/}}
\righthead{Two New Planets in Eccentric Orbits}
\slugcomment{To appear in the Astrophysical Journal, July 1999}

\received{}
\accepted{}

\begin{document}

\title{Two New Candidate Planets in Eccentric Orbits$^{1}$}

\author{Geoffrey W. Marcy\altaffilmark{2}, 
R. Paul Butler\altaffilmark{3},
Steven S. Vogt\altaffilmark{4}, 
Debra Fischer\altaffilmark{2}, 
Michael C. Liu\altaffilmark{5}}

\authoremail{paul@aaoepp.aao.gov.au}

\altaffiltext{1}{Based on observations obtained at the W.M. Keck
Observatory, which is operated jointly by the University of California
and the California Institute of Technology, and based on observations
obtained at Lick Observatory which is operated by the University of
California.}

\altaffiltext{2}{Department of Physics and Astronomy, San Francisco
State University, San Francisco, CA, USA 94132
and Department of Astronomy, University of California,
Berkeley, CA USA  94720}

\altaffiltext{3}{Anglo--Australian Observatory, PO Box 296, Epping, NSW, 1710, Australia}

\altaffiltext{4}{UCO/Lick Observatory, 
University of California at Santa Cruz, Santa Cruz, CA, USA 95064}

\altaffiltext{5}{Department of Astronomy, University of California,
Berkeley, CA USA  94720 }

\begin{abstract}

Doppler measurements of two G--type main--sequence stars, HD210277
and HD168443, reveal Keplerian variations that imply the presence of
companions with masses (\msini) of 1.28 and 5.04 \mjup and orbital
periods of 437 d and 58 d, respectively.  The orbits have large
eccentricities of $e$=0.45 and $e$=0.54, respectively.  All 9 known
extrasolar planet candidates with $a$=0.2--2.5 AU have orbital eccentricities
greater than 0.1, higher than that of Jupiter ($e$=0.05).  Eccentric orbits may
result from gravitational perturbations imposed by other orbiting
planets or stars, by passing stars in the dense star-forming cluster, or by
the protoplanetary disk.  Based on published studies and
our near-IR adaptive optics images, HD210277 appears
to be a single star.  However,
HD168443 exhibits a long--term velocity trend consistent with a 
close stellar companion, as yet undetected directly.

\end{abstract}

\keywords{planetary systems -- stars: individual (HD210277, HD168443, HD114762)}

\section{Introduction}
\label{intro}

Doppler surveys of main sequence stars have revealed 15 companions to
main sequence stars that are extrasolar planet candidates.  Among
these candidates, 13 have \msini $<$ 5 \mjup.  The host stars and
associated descriptions are: 51 Peg (Mayor \& Queloz 1995), 47 UMa
(Butler \& Marcy 1996), 70 Vir (Marcy and Butler 1996), 55 Cnc,
$\upsilon$ And, $\tau$ Boo (Butler et al. 1996), 16 Cygni B (Cochran
et al. 1997), $\rho$ CrB (Noyes et al. 1997), GJ876 (Marcy et
al. 1998, Delfosse et al. 1998), 14 Her (Mayor et al. 1998), HD187123
(Butler et al. 1998), HD195019A \& HD217107 (Fischer et al. 1999),
GJ86 (Queloz et al. 1999) and HD114762 (Latham et al. 1989).  The
Doppler measurements reported here suggest the presence of new
planetary candidates around HD210277 and HD168443.

Four main sequence stars harbor Doppler companions that
have \msini = 15--75 \mjup, which may represent the ``brown dwarfs''
(Mayor et al. 1998, Mayor et al. 1997).  Indeed the companions to 70
Vir (\msini = 6.8 \mjup) and to HD114762 (\msini = 11 \mjup) may also
represent ``brown dwarfs'' (Marcy \& Butler 1996, Mazeh et al. 1997,
Latham et al. 1989, Boss 1997).  The distinction between ``planets''
and ``brown dwarfs'' remains cloudy and rests on two formation
scenarios.  Planets form out of the agglomeration of condensible
material in a disk into a rock--ice core (eg. Lissauer 1995).  In contrast,
brown dwarfs presumably form by a gravitational instability in gas
(eg. Boss 1998, Burrows et al. 1998).  Hybrid formation scenarios
remain viable in which the relative importance of solid core growth
and gas accretion within a disk lead to a continuum in internal structure.
Subsequent collisions may lead to further growth and dynamical
evolution.  The current dichotomous taxonomy may describe substellar
physics no more precisely than ``spiral'' and ``elliptical'' summarize
galactic physics.

The first incontrovertible sub--classification within the
substellar regime is revealed in the mass distribution.  Companions
having \msini in the decade between 0.5--5 \mjup outnumber those
between 5--50 \mjup by a factor of $\sim$3 (eg, Marcy \& Butler 1998,
Mayor et al. 1998).  The poor detectability of the lowest--mass companions
implies that the factor of 3 is a lower--limit to the cosmic ratio.
This plentitude of companions having Jovian masses suggests that
qualitatively distinct formation processes predominated, arguably
similar to those associated with the giant planets in our Solar System
(Lin et al. 1998).

The extrasolar planets reveal some peculiarities that may bear on
their formation.  The host stars of the extrasolar planet candidates
have higher mean metallicity by a factor of $\sim$2 in abundance compared
with field stars (Gonzalez 1998).  Metallicity was not a criterion in
the selection of the target stars for these planet searches. Equally
interesting is that seven planets reside in orbits with a radius less than
0.12 AU, sometimes termed ``51 Peg'' planets (Mayor \& Queloz 1995,
Butler et al. 1998).  Precision Doppler surveys are most sensitive to
planets in small orbits, resulting in a selection effect.
Nonetheless, these small orbits challenge us to
explain their existence in a region where both the high temperatures
and the small amount of protostellar material would inhibit formation
{\it in situ}.  The 51 Peg planets thus offer support for the prediction
by Goldreich \& Tremaine (1980) and Lin (1986) that Jupiters may
migrate inward from farther out (Lin et al. 1995, Ward 1997, Trilling
et al.  1998).

Perhaps most intriguing about the planet candidates are the orbital
eccentricities.  The orbits of the 51 Peg planets may suffer some
tidal circularization (Lin et al. 1998, Terquem et al.1998, Ford et
al. 1998, Marcy et al. 1997), and indeed their orbits are all nearly
circular (see Table 4).  In contrast, the planets that orbit farther
than 0.15 AU from their star all reside in non--circular orbits having
$e>$0.1, i.e. more eccentric than for Jupiter ($e$=0.05).  Indeed, all
but two have $e>$0.2 .  This high occurrence of orbital eccentricity
has lead to a variety of models in which Jupiter--like planets suffer
gravitational interactions with a) other planets (Weidenschilling \&
Marzari 1996, Rasio \& Ford 1996, Lin \& Ida 1996, Levison et
al. 1998), b) the disk (Artymowicz 1993), c) a companion star (Holman
et al. 1996, Mazeh et al. 1997), and d) passing stars in the young
cluster (de la Fuente Marcos and de la Fuente Marcos 1997, Laughlin
and Adams 1998).

The possibility persists that the observed non--circular orbits all
stem from perturbations from a bound companion star, as proposed for
16 Cyg B (Holman et al. 1997, Mazeh et al 1997), rather than from
intrinsic dynamics of planet formation.  This paper reports the
detection of two new planetary candidates orbiting at 0.3 and 1.1 AU,
both having large eccentricities.  The observations and orbital
solutions are reported in section 2.  The search for stellar
companions is discussed in section 3.  Section 4 contains a discussion
of the implications for planet formation.

\section{Observations and Orbital Solutions}
\label{obs}

\subsection{Stellar Characteristics of HD210277}

The two stars described here are among 430 G,K, and M-type main
sequence stars currently being monitored at the Keck I telescope for
Doppler variations.  HD210277 has an effective temperature of 5570
$\pm$50 K, the average determination from spectral synthesis of
high--resolution spectra (Favata et al. 1997, Fuhrmann 1998, Gonzalez
et al. 1998), which also yields a surface gravity of log $g$ =
4.38 $\pm$ 0.1 (Fuhrmann 1998, Gonzalez et al. 1998).  These
surface values imply main--sequence status and correspond to spectral
type G7V (Gray 1997).

The metallicity of HD210277 is measured to be [Fe/H] = +0.24
$\pm$0.02, considerably higher than the average value for field stars,
$<$[Fe/H]$>$ = --0.23, in the Solar neighborhood (Gonzalez et al. 1998, Favata et al. 1997, Fuhrmann 1998).  Thus, HD210277
appears to be rich by a factor of 3 in its abundance of heavy
elements, normalized to hydrogen, placing its metallicity within
the upper 5\% of nearby stars.  We measure a radial velocity of --21.1
$\pm$ 2 \kms, which agrees with that of Duquennoy and Mayor (1991),
--21.44 \kms.  Its parallax of 0.047 arcsec (Perryman et al. 1997)
implies an absolute visual magnitude of M$_V$=4.90$\pm$0.05 and a
luminosity, $L$=0.93 L$_{\odot}$.  These stellar parameters permit
placement of HD210277 on evolutionary tracks, which yield a mass,
$M$=0.92 $\pm$ 0.02 \msun and an age of 12 $\pm$2 Gyr (Gonzalez et al. 1998).

One stellar characteristic that bears on the Doppler detectability of
planets is the magnetic field and chromosphere.  Spots on a
rotating star can produce spurious Doppler shifts, and chromospheric
emission correlates with spurious Doppler ``noise'' presumably caused
by surface magnetohydrodynamics (Saar et al. 1998).  Our spectra
contain the chromospheric H\&K emission lines from which stellar
rotation and stellar age can be estimated (Noyes et al. 1984).  This
emission yields the chromospheric index known as the ``Mt Wilson S
Value'' of $S$ = 0.155, implying $R$'(HK)=-5.06, measured from 36
spectra obtained from 1996.5 through 1998.7 (Shirts \& Marcy 1998).
See Baliunas et al. (1998) for a detailed discussion of the $S$ value.
No trend or periodicity are apparent in the S values of HD210277, and
the RMS is 0.006, all of which indicate that HD210277 is
chromospherically quiet.  The implied rotation period is, $P_{\rm
Rot}$=40.8 d and the age is 6.9 Gyr.  In conjunction with the
aforementioned age of 12 Gyr from tracks, we conclude that HD210277
has an age in the range 7--10 Gyr, but not evolved into the subgiant
regime.  Such a chromospherically inactive star may produce spurious
Doppler shifts of no more than $\sim$3 \ms (Saar et al. 1998, Butler
et al. 1998).

\subsection{Stellar Characteristics of HD168443}

No detailed LTE analysis of HD168443 has been carried out to our
knowledge.  A photometric analysis was done by Carney et al. (1994)
giving T$_{\rm eff}$= 5430 and $m/H$=--0.14.  The metallicity is
apparently slightly subsolar, similar to the mean for nearby field
stars.  Its parallax of 26.4 mas (Perryman et al. 1997) implies an
absolute visual magnitude of M$_V$=4.03$\pm$0.07 and a luminosity,
$L$=2.1 L$_{\odot}$, which places it 
$\sim$1.5 mag above the main sequence at its
T$_{\rm eff}$.  These stellar parameters suggest a subgiant status and
spectral type, G8IV.  Apparently, HD168443 is similar to 70 Vir in
mass, surface characteristics, and metallicity (Marcy \& Butler 1996,
Apps 1998).

Our spectra of HD168443 yield a chromospheric S Value of $S$ = 0.147,
with an RMS of 0.009 during 30 observations from 1996--1998.5,
implying $R$'(HK)=-5.08.  No trend or periodicity are apparent in the
S values.  The implied rotation period is $P_{\rm Rot}$=37 d, and the
implied age is 7.8 Gyr, from the calibration by Noyes et al. (1984).
In conjunction with its possible subgiant status from above, we
conclude that HD168443 has an age of 7--10 Gyr, slightly evolved
toward subgiant status.  We caution that the subgiant status remains
in question, pending spectroscopic assessment of surface gravity.

A mass determination for HD168443 is given by Carney et al. (1994) who
find $M$=0.84 \msun.  This mass determination may warrant revision
because it preceded the Hipparcos parallax and because it did not
include revisions to the metallicity dependence of evolutionary tracks
(Bertelli et al. 1994).  Based on the Hipparcos data and new tracks,
along with available narrow--band photometry for HD168443, Apps (1998)
estimates a mass of 1.05 $\pm$0.10 \msun for HD168443.  We adopt here
the straight average of the two mass estimates to yield $M$ = 0.945
$\pm$0.1 \msun .

We measure a radial velocity of --49.0 $\pm$ 2 \kms (on 1998 Aug),
which along with its high transverse velocity of 44 \kms, suggests a
kinematic association with the old disk population.  Such an old,
chromospherically inactive star may produce spurious Doppler shifts
$\sim$3 \ms of photospheric origin (Saar et al. 1998, Butler et
al. 1998).

\subsection{Details of the Doppler Measurements}

For both HD210277 and HD168443, spectra were obtained from 1996.5
through 1998.7 with the HIRES echelle spectrometer on the Keck I
telescope (Vogt et al. 1994).  We used slit ``B1'' that has a width of
0.57 arcsec and height of 3.5 arcsec.  The resolution for these
spectra was $R$=87000, based on the measured FWHM of the spectrometer
instrumental profile.  The spectra span wavelengths from 3900--6200
\AA.  The wavelength scale and instrumental profile were determined
for each 2--{\AA} chunk of spectrum for each exposure by using iodine
absorption lines superimposed on the stellar spectrum (Butler et
al. 1996).  The measured velocities are relative, with an
arbitrary zero--point.

The typical exposure times were $\sim$5 minutes, depending on seeing,
for both stars, yielding a S/N=300 per pixel (1/2 of one resolution
element).  Such spectra are expected to carry photon--limited Doppler
precision of 2--3 \ms (Butler et al. 1996).  Indeed, the uncertainty
in the mean velocity of the 400 spectral chunks is typically 2.5 \ms.
However, our results from 430 stars on the survey reveal a median RMS
velocity of 6 \ms, which we interpret as the actual scatter that
limits planet detection.  Intrinsic photospheric noise of $\sim$3 \ms
accounts for some of the 6 \ms scatter (Saar et al. 1998).  This
intrinsic stellar effect may be added in quadrature to the
photon--limited errors of 2.5 \ms to establish an expected Doppler
scatter of 3.9 \ms.  Thus, we infer that unidentified errors of
$\sim$4 \ms persist in our Doppler results which presumably stem from
inadequacies in our spectral modelling, improvements to which are in
progress.

\subsection{Keplerian Velocities for HD210277}

The 34 measured velocities of HD210277 are listed in Table 1 along
with the JD date.  Again, the true uncertainty of each measurement is
$\sim$6 \ms.  A plot of the velocities for HD210277 is shown in
Figure 1.  It is apparent that the velocities for HD210277 scatter
with a peak--to--peak variation of $\sim$80 \ms, and the velocities
are correlated in time.  A periodogram analysis revealed no
significant peak because too few cycles have transpired
during the two years of observations and because a Lomb-Scargle
periodogram is not robust for non-sinusoidal variations which result from 
eccentric orbits.

A suggestive period of 1.2 yr is evident in the velocities for HD210277, though less than two periods have transpired.  The best--fit
Keplerian model yields an orbital period of 437$\pm$25 d, a
semi-amplitude $K$ = 41.0$\pm$5 \ms, and an eccentricity $e$ =
0.45$\pm$0.08 .  The complete set of orbital parameters are given in
Table 3.  The RMS to the Keplerian fit is 7.1 \ms, similar to the
uncertainty and similar to the velocity RMS, 7.6 \ms, for the orbital
fit to a previously-discovered Keck survey planet HD187123 (Butler et
al. 1998). Thus, the RMS of 7 \ms for HD210277 implies that a single
companion provides a model that plausibly explains the velocities.

Using the stellar mass of 0.92 \msun, the companion mass is
constrained as, \msini = 1.28 \mjup , and the semimajor axis is $a$ =
1.10 AU.  With periastron and apastron distances of 0.61 and 1.60 AU,
HD210277 is unlikely to harbor additional companions within that
range.

\subsection{Keplerian Velocities for HD168443}

The 30 velocity measurements for HD168443 are listed in Table 2 along
with the JD date. As with HD210277, the true uncertainty of each
measurement is $\sim$6 \ms.  A plot of the velocities for HD168443 is
shown in Figure 2. The velocities for HD168443 scatter with a
peak--to--peak variation of 650 \ms, with clear temporal correlations
and trends among measurements. A periodogram reveals two dominant
peaks, at $P$=20 d and $P\approx$55 d.

We carried out nonlinear least--squares fits of Keplerian models to
the velocities, starting with trial periods ranging from 3 -- 600 d.
For trial periods near 20 d, the lowest velocity RMS was 69 \ms which
is clearly inconsistent with expected scatter of 6 \ms.  For trial
periods near 55 d, we found two nearby minima in $\chi^2$,
corresponding to two slightly different orbital periods, $P$=64 d
(RMS=23 \ms) and $P$=59 d (RMS=36 \ms).  Figure 3 shows the
velocities as a function of orbital phase for the better of those fits
($P$=64 d).  That Keplerian fit carries an implied eccentricity of
$e$=0.69, $K$=292 \ms, and a companion minimum mass of \msini = 4.0
\mjup.

However the scatter to that fit, RMS=23.3 \ms , clearly exceeds the
expected scatter of 6 \ms , implying that this fit carries a reduced
$\chi^2$ greater than 4 and hence this model is inadequate.  Indeed,
two telltale points located at phase $\sim$0.95 (see Figure 3) were obtained
on consecutive nights.  The second velocity was 58 \ms higher, and yet
according to the Keplerian curve, it should reside lower by 60 \ms.
We consider this dubious orbital fit to imply that the Keplerian model
fails in some important way.

We modified the model by simply adding a variable linear trend to
Keplerian velocities.  Such a model incorporates the possibility of a
long period companion in addition to the shorter--period companion.
This slope introduces only one additional free parameter, as the
``y-intercept'' of the slope is subsumed within the arbitrary
zero--point of the velocities.

The Keplerian--plus--slope model is shown in Figure 4 and 
yields a best fit orbital period of,
$P$=57.8 d, $K$=350 \ms, $e$=0.54, and \msini=5.04 \mjup .  The RMS of
the residuals, 12.8 \ms, is considerably reduced from RMS=23.3 \ms that
results from a model without a trend.  The reduced $\chi^2$ for this
solution is 2.3.  All orbital parameters are given in Table 3.  Thus,
it appears that the introduction of an {\it ad hoc} slope into the
Keplerian model for HD168443 significantly improves the fit.
However, the RMS of 12.8 \ms remains larger than the expected scatter
of 6 \ms, implying that the addition of a velocity slope is too
simple.  Introducing an {\it ad hoc} parabolic term in the velocity
trend reduces the RMS to 8 \ms ($\chi^2$=1.5), superior to that of a
linear trend.  However, we feel that introducing this parabolic free
parameter carries only marginal statistical justification.  A proper
model that contained a second orbiting companion would require the
introduction of an additional set of Keplerian parameters, for which
we have inadequate constraints. Thus, the only model supported by the
current data is that containing the linear trend.

We tested the predictability of this model with two additional
velocity measurements obtained with the 0.6--m CAT telescope at Lick
Observatory.  We obtained spectra on two consecutive nights, centered
on Julian Dates 2451100.644 and 2451101.645 for which the model
containing the Keplerian and linear trend offered a prediction of an
increase in velocity of 52.4 \ms.  On both nights we obtained four
consecutive spectra, each lasting 30 min.  Each spectrum was analyzed
separately to derive a Doppler shift.  The four velocities were
averaged, to yield the final velocity for each night.  The uncertainty
in the mean was computed from the standard deviation of the four
separate measurements, giving an internal error for each night.

These two Lick velocities were --24.2 $\pm$ 6.2 \ms and +24.1 $\pm$ 3
\ms on the two nights, respectively, implying that the velocity of HD168443 
increased by +48.3 $\pm$ 7 \ms.  This velocity increase agrees
with the prediction of the model (Keplerian plus trend) of +52.4 \ms.
These Lick velocities were obtained 26 days after the last Keck
measurements were made, on JD=24511074.785, shown in Figure 4.  The
alternative model without an imposed velocity trend has a longer
period of $P$=64.3 d, and its predicted change in velocity is --60
\ms, clearly in conflict (wrong sign) with the observed rise of 48
\ms.  Thus, both the lower velocity RMS and the Lick measurements
favor the model that contains a Keplerian with $P$=57.8 d and a
velocity trend. 

The best--fit velocity trend has slope of 89.4 \ms per yr which could
be caused by a second more distant companion.  If so, its minimum
orbital period is $\sim$4 yr.  As a benchmark, Jupiter causes a trend
of $\sim$4 \ms per yr in the Sun during 6 yr.  Thus, if the period of
the hypothetical second companion to HD168443 were $\sim$12 yr, its
mass would be at least $\sim$25 \mjup.  For the shortest possible
period of 4 yr, the companion mass would be at least 15 \mjup.  In
both cases, the companion would be considered a ``brown dwarf'' and
quite possibly a hydrogen--burning star, depending on the actual
period and $\sin i$.  Prospective stellar companions are discussed in
section 3.2

\subsection{Velocities for HD114762}

We have obtained 33 velocity measurements for HD114762 since 1994
Nov.  They are plotted versus orbital phase in Figure 5.  The unseen
companion to this star has been described by Latham et al. (1989),
Mazeh et al. (1996), Cochran et al.  (1991), and Hale (1995).  Our
velocities offer new measurements of the orbital parameters,
$P$=84.03$\pm$0.1 d, $e$=0.334$\pm$0.02, $K$=618$\pm$6 \ms,
$\omega$=201$\pm$3 deg, and $T_p$=JD2450225.30$\pm$0.6.  A revised
mass for HD114762 has been measured by Ng and Bertelli (1998) and
Gonzalez (1998), giving, $M$=0.82$\pm$0.03 \msun, based on its
Hipparcos distance ($d$=40.57 pc, Perryman et al. 1997) and new
stellar evolution models.  This stellar mass and the orbital
parameters imply that the companion has \msini = 11.02$\pm$0.5 \mjup.

If the companion mass is truly small compared to the primary star then
the semimajor axis is $a$=0.35 AU.  However, Cochran et al. (1991) and
Hale (1995) provide arguments that the companion mass may be large,
possibly stellar.  Since that work, several additional considerations
have emerged regarding its status as a candidate planet.  HD114762 is
the only planet-candidate found with modest velocity precision, rather
than with high precision of $\sim$10 \ms.  That precision along with
the large survey size (Latham et al. 1989) makes the discovery of an
face-on system more likely.  Further, HD114762 has [Fe/H]=-0.6, substantially 
more metal--poor than any other planet candidate (Gonzalez 1998).  The
standard model of planet formation requires heavy elements to form the
dust which was presumably not abundant in the protoplanetary disk
around HD114762.  Finally, the value of \msini (11.02 \mjup) is much
higher than that for all other planet candidates, the highest of which
is \msini = 7.4 \mjup (70 Vir).  Nonetheless, we include HD114762 as a
candidate planetary object in this complete compilation.

\section{Search for Stellar Companions}

\subsection{HD210277}

We examined HD210277 for companion stars as follows.  Duquennoy and
Mayor (1991) made 8 radial velocity measurements spanning 6 yr which
exhibited no variation above 220 \ms.  Ground--based astrometry from
1989--1993 revealed no motion at a level of 0.01 arcsec (Heintz 1994).
Lunar occultation measurements revealed no companion to HD210277,
with detection thresholds of $\Delta$ V$_{\rm mag}<$2 within 3 mas
Meyer et al. (1995) .  The above measurements, especially those of
Duquennoy and Mayor, jointly rule out stellar companions with masses
as low as 0.1 \msun within 10 AU.  A 0.1 \msun$~$ dwarf orbiting 10
AU from HD210277 would induce velocity variations with semi-amplitude
of 700 \ms ($\times \sin i$) and a period of 30 yr, detectable as a
trend in velocities of Duquennoy and Mayor (1991), but not observed.
The astrometry of Heintz similarly rules out a stellar companion with
mass down to the substellar limit within 5 AU, which would have induced
astrometric wobbles of 0.02 arcsec.

To search for possible stellar companions beyond 10 AU, we observed 
HD210277 on 8 September 1998 UT using the Lawrence Livermore National
Lab adaptive optics system (Max et al. 1997), which is mounted at the
f/17 Cassegrain focus of the Lick Observatory Shane 3-m telescope.
The adaptive optics system performs real-time compensation of
atmospheric seeing using a Shack-Hartmann type wavefront sensor with a
127-actuator deformable mirror.  In its current configuration, 61 of
the actuators are actively controlled.  For these observations, image
compensation was done with a sampling frequency of 250~Hz using 
HD210277 (V=6.54) itself as a wavefront reference, achieving a
closed-loop bandwidth of 20 Hz.

We acquired images using the Lick facility near-IR camera LIRC2
(Gilmore, Rank, \& Temi 1994).  The camera has a 256~$\times$~256
pixel HgCdTe NICMOS-3 detector and, when coupled with the AO system, a
plate scale of 0.12 arcsec/pixel.  We used both a narrow-band
($\Delta\lambda/\lambda = 0.01$) filter centered on Br$\gamma$ (2.166
\micron) and the broad-band K$^\prime$ filter (1.95-2.35 \micron;
Wainscoat \& Cowie 1992) to span a wide range in radii with good
sensitivity and dynamic range. The star was dithered to 4 positions on
the detector, with total integrations of 240~s in each filter.  Images
were reduced in a standard fashion for near-IR images --- bias
subtraction, flat-fielding, and sky subtraction using a master sky
frame constructed from all the images. The angular resolution as
measured by the full-width at half-maximum (FWHM) of the Br$\gamma$
images is 0.18 arcsec, and the images have a mean Strehl ratio of
0.45.  The Br$\gamma$ data are most sensitive to companions inside of 0.5
arcsec, and the $K^\prime$-images are more sensitive at larger radii.

Figure 6 presents our 4$\sigma$ upper limits to any stellar companions
to HD210277 combined with $K$-band flux ratios for main sequence
companions derived from Kirkpatrick \& McCarthy (1994).  Only the
inner radii are shown for clarity; the deepest portion of the our
images cover 12 arcsec in radius. Our AO data are nearly
diffraction-limited, ruling out any main-sequence dwarf companions
earlier than spectral type M0 from 0.2--12 arcsec (4.2--250 AU) .  In
addition, the high Strehl ratio means the images are very sharply
peaked and sensitive to even the lowest mass M dwarfs outside of 0.5
arcsec (11 AU).  We rule out any main sequence companion with a
separation of 0.8 arcsec (17 AU) to 12 arcsec (250 AU).

We further rule out any stellar companions out to $\approx1\arcmin$
separations using 2.5 arcsec FWHM $J$ and $K^\prime$ images obtained
from the Lick 3-m with the UCLA two-channel infrared camera known as
"Gemini" (McLean et al. 1994).  The $J$ and $K^\prime$ data were taken
simultaneously on 09 October 1998 UT with two 256~$\times$~256
detectors, a Rockwell HgCdTe NICMOS-3 one for $J$ and a Hughes-SBRC
InSb one for $K^\prime$.  There are a handful of $K\approx 14-17$
unresolved sources in these images; the majority of these also appear
on the Palomar Sky Survey.  Their $J-K$ colors are consistent with
background stars or galaxies, and their numbers are in accord with
field $K$-band galaxy counts (Szokoly et al. 1998).  Comparing our
images with those on the Palomar Sky Survey, the only source which
shows noticeable proper motion is HD210277 itself, which exhibits 
a magnitude and direction consistent with its nominal proper motion.

\subsection{HD168443}

We have searched for stellar companions to HD168443 in several ways.
A literature search turned up no known companions.  We searched for
superimposed spectral lines from a secondary star in the 8400 \AA\
region of our Lick spectra (near the CaII IR triplet).  No such lines
were found at a threshold of a few percent of the continuum.  This
nondetection rules out any main--sequence companions more massive than
0.5 \msun within 2.5 arcsec (95 AU) of HD168443, as we use a slit
width of 5 arcsec with the Lick Observatory Coud\'e Auxiliary
Telescope.

Carney et al. (1994) obtained 8 radial velocity velocity measurements
of HD168443 spanning 5 yr and detected no variation above the errors
of 400 \ms. Any stellar companion more massive than 0.1 \msun$~$
orbiting within 5 AU would have been revealed, except for extreme
values of $\sin i$.  The Hipparcos astrometry of HD168443 recorded no
astrometric motion at a level of 2 mas during several years (Perryman
et al. 1997).  A stellar companion having 0.1 \msun\ at 5 AU would
induce a (curved) astrometric reflex motion of 16 mas during $\sim$3
yr (1/4 orbital period), as viewed from a distance of 38 pc.  Such a
wobble evidently did not occur, thus ruling out stellar companions
within 5 AU, consistent with the velocity data.  Hipparcos would not
easily detect stellar companions orbiting beyond 5 AU, as the (more
linear) reflex motion could be absorbed into the assessment of proper
motion.  Thus, stellar companions orbiting beyond 5 AU might escape
detection by both the Carney et al. velocities and the Hipparcos
astrometry.

We have not obtained an adaptive optics image of the star, leaving us
little information about stellar companions farther than 5 AU.
However, we were compelled to include a velocity slope of 89 \ms per
yr in the model of our velocities (Fig 4).  This slope could indicate
a stellar companion beyond 5 AU, or a brown dwarf somewhat closer.  As
a benchmark, a 0.1 \msun$~$ companion orbiting at 10 AU would induce a
typical velocity slope of 100 \ms per yr ($\times \sin i$).  Our Keck
velocities are consistent with such a stellar companion as well as
more distant and correspondingly more massive ones.  The upper mass
limit of 0.5 \msun, imposed by the lack of secondary lines, implies
that the companion must reside within $\sim$30 AU in order to induce
the observed velocity slope of 89 \ms per yr.

A consistency check on the putative companion is provided by comparing
the absolute velocities obtained by Carney et al (1994) to those found
here.  Carney et al. obtained 8 velocity measurements centered at
epoch $\sim$1990 which exhibited an average of --48.9 \kms, with $\sigma$=0.4
\kms.  Our observation on 1998 Aug 25 gave a velocity of --49.0$\pm$2
\kms , which agrees with the Carney measurement within the 2 \kms
uncertainty.  This implies an upper limit to the velocity trend of 2
\kms per 8 yr, which is indeed larger than the trend we actually
detect of 89 \ms per year.

In summary, any stellar companion must reside beyond 5 AU but not
beyond 30 AU to explain the observed velocity trend, and its mass must
be less than 0.5 \msun to explain the lack of stellar secondary lines.
A direct search for a stellar companion located 0.15--1 arcsec from HD
168443 seems warranted.

\section{Discussion}

The two extrasolar planet candidates suggested by the data in this paper bring
the total number of such candidates to 17.  These candidates all have
\msini $\la$ 5 \mjup except 70 Vir (\msini = 7.4 \mjup) and HD114762
(\msini = 11 \mjup)
which some would place in the ``brown dwarf'' class (Black 1998).  
Table 4 lists the basic orbital
parameters and \msini of all 17 known planetary candidates.
A few of the orbital parameters have been slightly modified,
based on our own recent measurements and orbital fits.
The typical uncertainty in the orbital 
eccentricity is 0.03, based on Monte Carlo
simulations of the Keplerian fits to data with artificial noise

Table 4 shows that all 9 planet candidates that have a$>$0.2 AU have
eccentricities above 0.1, larger than that for both Jupiter ($e$=0.048) and
Saturn ($e=$0.055) .  Figure 7 shows a plot of orbital
eccentricities vs. semimajor axis.  All extrasolar planets orbiting
closer than 0.1 AU have small eccentricities.  While possibly
primordial, these near--circular orbits for close planets may have
been induced by tidal circularization (cf., Rasio et al. 1996, Marcy
et al. 1997, Terquem et al. 1998).

Apparently, Jupiter--mass companions orbiting from 0.2--2.5 AU, immune
to tides, have large orbital eccentricities.
Apparently, some mechanism commonly produces eccentric orbits
in Jupiter--mass companions that reside from 0.2 -- 2.5 AU in
main--sequence stars.  These eccentric planets represent a general
property of 0.3--1.2 \msun\ stars .

Figure 8 shows orbital eccentricities vs. \msini.  No trend is
apparent at first glance, suggesting that orbital eccentricity is not
correlated with planet mass, within the mass range 0.5--5 \mjup.
However, all 5 planets with \msini$<$1.1 \mjup reside in nearly
circular orbits. This correlation may be a selection effect, as the
lowest--mass planets are more easily detected close to their host
stars in order to induce a detectable Doppler reflex signal.  These
close planets are all subject to tidal circularization.  Thus, the low
eccentricities among the lowest mass giant planets may not be
considered intrinsic to planet formation.

Figure 9 shows \msini vs. semimajor axis for all 17 planet candidates.
The detectability of planetary companions is shown as the curved line
near the bottom (Cumming et al. 1999).
Apparently, the distribution of planet masses is not a strong function
of semimajor axis from 0.05--2.5 AU for the range of detectable masses, 1--6
\mjup.  There is no paucity of either the most or least
massive companions at either extreme of semimajor axis.  Of course
there may be some blurring in mass due to $\sin i$.  Nonetheless, 
we conclude
that if orbital migration within a gaseous disk brings the giant
plants inward, neither that process nor the halting mechanism seems to
depend on planet mass.

Figure 10 shows a histogram of Msini within the range
0$<$15 \mjup for known companions to main--sequence stars.  The
distribution of \msini shows a rapid decline at
roughly 4 \mjup.  There are no companions having \msini = 7.5--11 \mjup
and those massive companions would have been easily detected.  This
absence seems statistically significant relative to the 14
companions having \msini =0.4--5 \mjup. All selection effects favor
detection of the high--mass companions, and thus the apparent drop in
the \msini histogram from 4 to 7 \mjup must be real.  This drop implies that
the distribution of companion masses, d$N$/d$M$, must indeed 
exhibit a decline  at $\sim$5 \mjup with increasing mass, within 2.5 AU.

The highest value of \msini among planet candidates (Fig 10) is for
HD114762 which has \msini=11.02 \mjup, which is well above the decline
at $\sim$5 \mjup.  Its unknown $\sin i$ leaves an important question
unanswered regarding its true mass and hence any affiliation with
``planets.''  With that possible exception of HD114762, the planetary
mass distribution certainly declines rapidly for masses above 5 \mjup.

The origin of the distribution of semimajor axes and eccentricities
now presents a puzzle.  In the standard paradigm, giant planets form
outside 4 AU (Boss 1995, Lissauer 1995).  Inward orbital migration
(Lin et al. 1995, Trilling et al. 1998) within the gaseous
protoplanetary disk has been suggested to explain the small orbits
detected to date among extrasolar planets.  Such migration makes two
predictions that appear testable.  First, orbital migration in a
viscous, gaseous environment is expected to preserve circular orbits
under most circumstances (but see Artymowicz 1993).  In contrast, all
9 planet candidates orbiting between 0.2 and 2.5 AU have non--circular
orbits.  Second, the orbital migration time scale is proportional to
the orbital period, which leads to rapid orbital decay for
successively smaller orbits.  In contrast, the observed orbital
semimajor axes are spread throughout 0.1 to 2.5 AU (though not
necessarily distributed uniformly).  No obvious mechanism is known to
halt the migration for these orbits.

Apparently, the orbits with sizes of $\sim$1 AU and large
eccentricities ($e>$0.1) require physical processes that are not explicitly
included within the context of quiescent migration in a dissipative medium.  
Scattering of orbits by
other planets, companion stars, or passing stars in the young star
cluster offer mechanisms for producing eccentric orbits (Rasio and
Ford 1995, Lin and Ida 1996, Weidenschilling and Marzari 1996,
Laughlin and Adams 1998).  However, these mechanisms do not explicitly
predict small orbits of $\la$1 AU, because significant
energy must be lost from the original orbits of $\sim$5 AU.

One possibility is that planet--scattering continues to occur during
the final era of the remnant gaseous protoplanetary disk.  If the disk
remains intact within the inner few AU where the original gas density
was highest, the disk can serve as the reservoir into which the
planet's orbital energy can be deposited, either by dynamical friction
or by tidal interaction between planet and disk.  In this scenario,
scattered planets would reside in eccentric orbits subjecting them to
dissipation during periastron passages.  Clearly detailed models are
required that include both planet scattering and the dissipative
effects of a weak inner gaseous disk to determine the resulting
planetary orbits.

In any case, we currently have little information about giant planets
that orbit beyond 3 AU.  We expect to obtain such information in the coming
years as Doppler programs extend their time baseline.   Planets beyond 3 AU 
may well reside in predominantly
circular orbits.  A population of giant planets that never suffered
significant scattering or migration could comprise these Jupiter
analogs.  The lack of main sequence stars having reflex Doppler
periodicities with amplitudes above 30 \ms already indicates a 
paucity of planets having $M>$3 \mjup within 5 AU (Cumming et
al. 1999).  It remains to be determined if the planet mass function
rises rapidly for smaller masses.

\acknowledgements

We thank Kevin Apps for assessment of stellar characteristics, Phil
Shirts for his measurement of Ca II H\&K, and Eric Williams for work
on HD114762.  We thank Claire Max, Scot Olivier, Don Gavel and Bruce
Macintosh for the development of the Lick adaptive optics system and
for assistance with the observations.  We thank Tom Bida, Tony Misch
and Wayne Earthman for technical help with instrumentation.  We
acknowledge support by NASA grant NAGW-3182 and NSF grant AST95-20443
(to GWM), and by NSF grant AST-9619418 and NASA grant NAG5-4445 (to
SSV), and by Sun Microsystems.  We thank the NASA and UC telescope
assignment committees for allocations of telescope time.

\clearpage
\begin{figure}
\psfig{figure=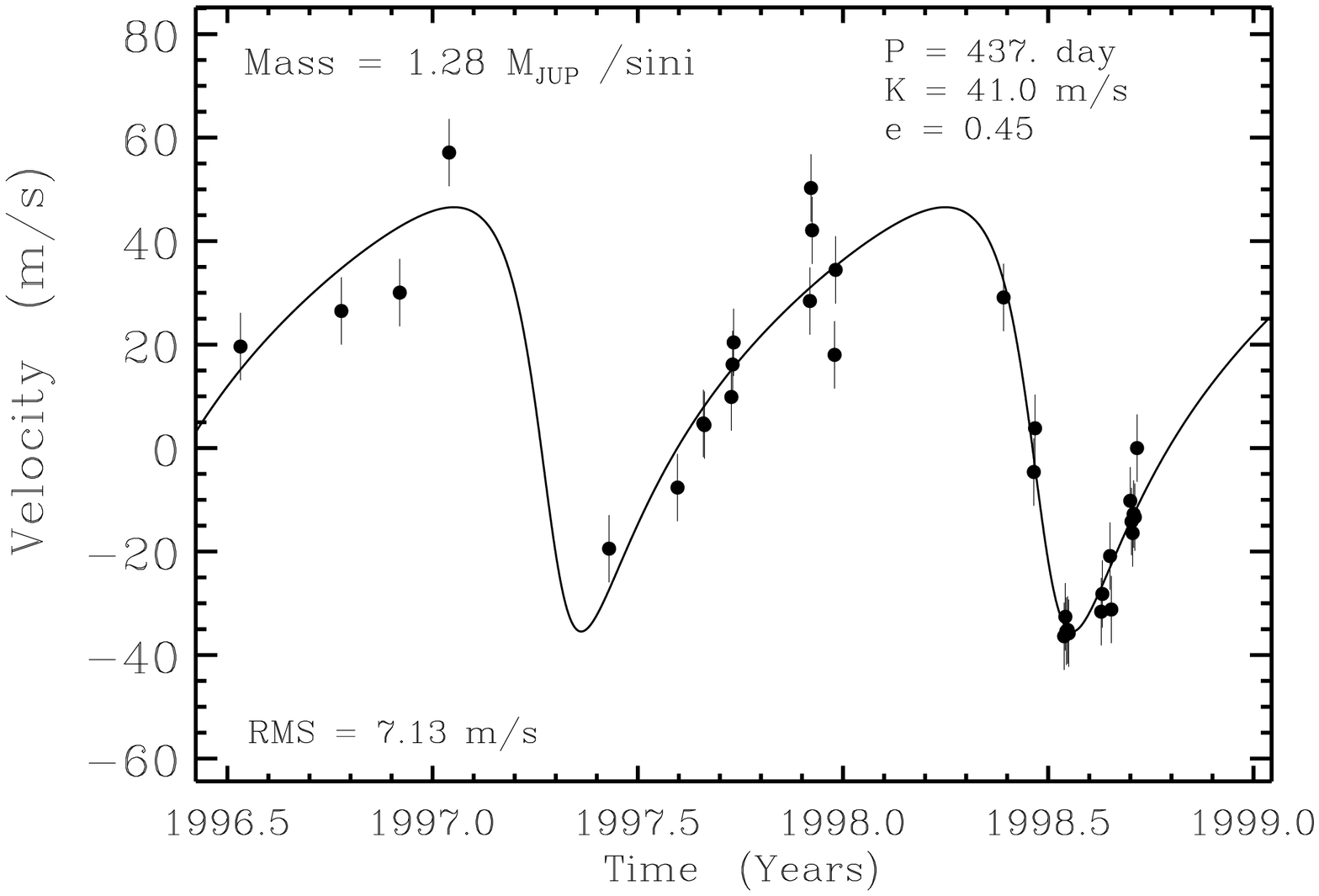,angle=90,width=6.5in}
\caption{Measured radial velocities for HD210277.  The solid line shows the
best--fit Keplerian curve.}

\label{rv_curve}
\end{figure}

\begin{figure}
\psfig{figure=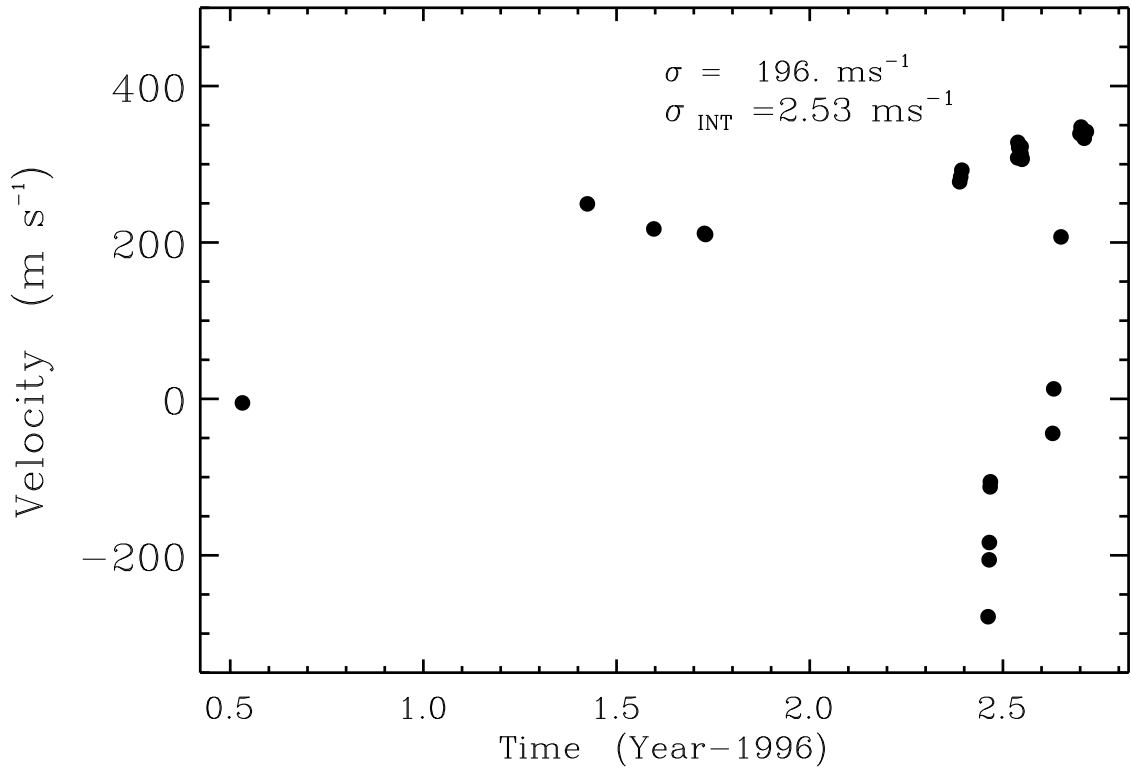,angle=90,width=6.5in}
\caption{Measured radial velocities for HD168443.  The points exhibit obvious
correlations in time, with a hint of periodicity.}
\label{fig2}
\end{figure}

\begin{figure}
\psfig{figure=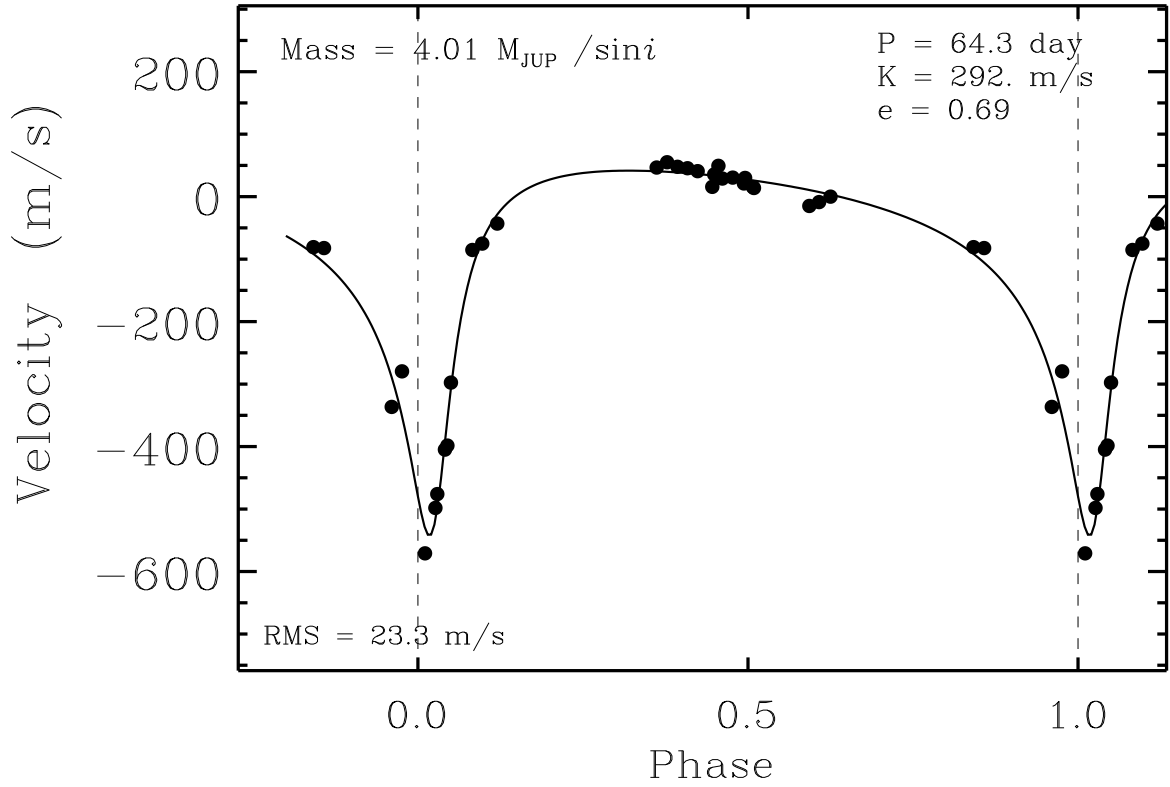,angle=90,width=6.5in}
\caption{The velocities of HD168443, plotted as a function of orbital
phase for the best--fit Keplerian model (without an {\it ad hoc}
velocity trend). This orbit has RMS residuals of 23.3 \ms, exceeding
the expected scatter by a factor of 4. This model appears inadequate
to explain the velocities.}
\label{fig3}
\end{figure}

\begin{figure}
\psfig{figure=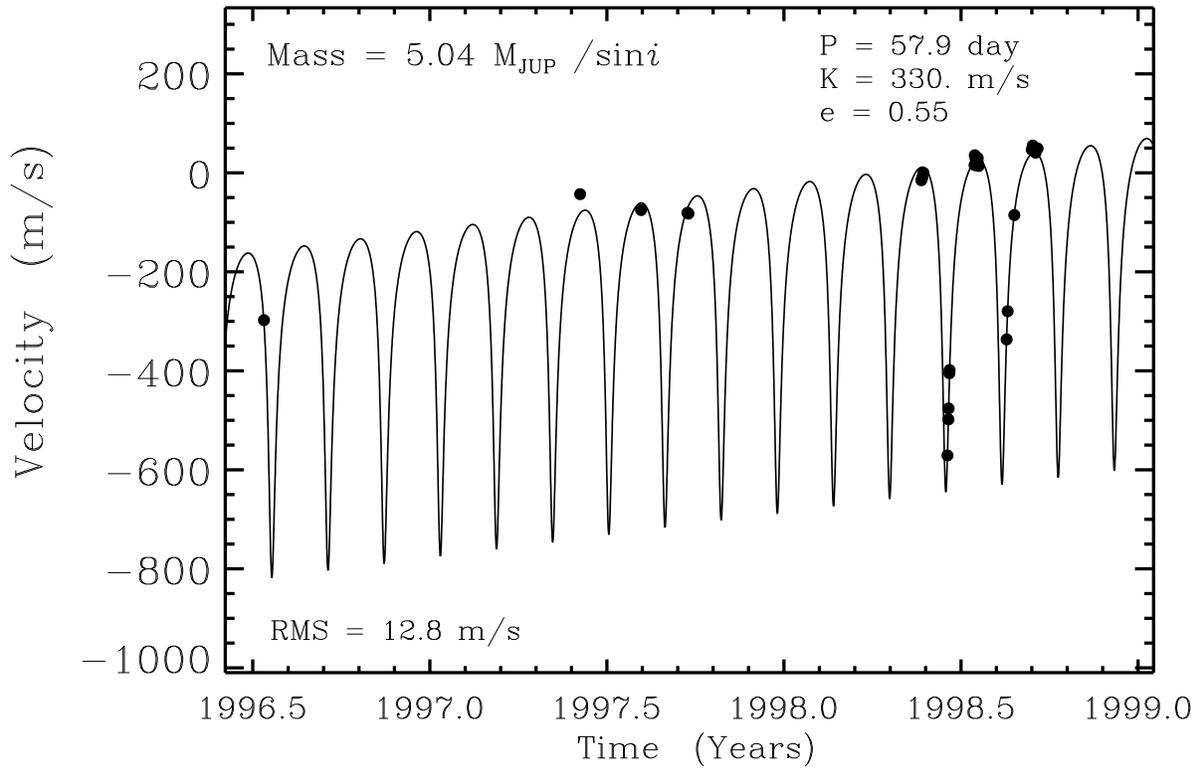,angle=90,width=6.5in}
\caption{The velocities of HD168443 plotted versus time.  The solid
line shows the best--fit Keplerian model with an added linear 
trend in velocity.
This {\it ad hoc} model yields residuals with RMS=13 \ms which is
about twice the expected scatter, but a clear improvement over the model
without a trend in Fig. 3.}
\label{fig4}
\end{figure}

\begin{figure}
\psfig{figure=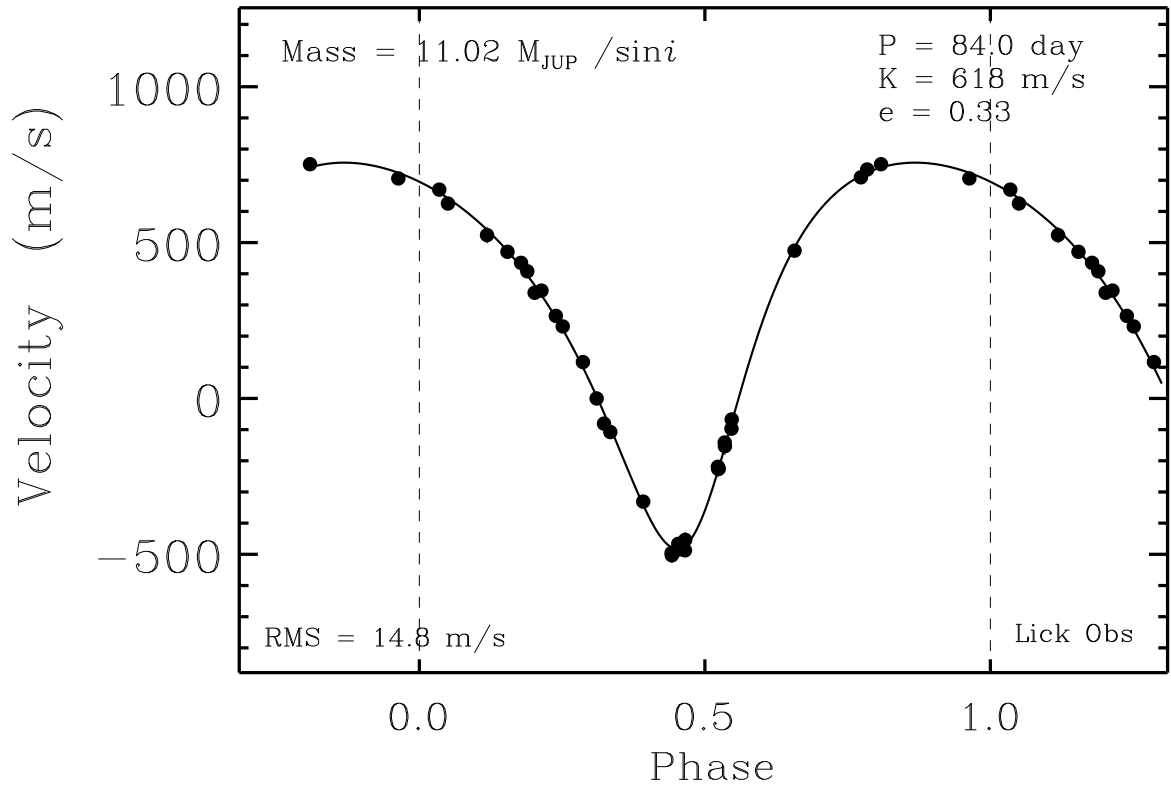,angle=90,width=6.5in}
\caption{The phased radial velocities of HD114762 from Lick Observatory.}
\label{fig5}
\end{figure}

\begin{figure}
\psfig{figure=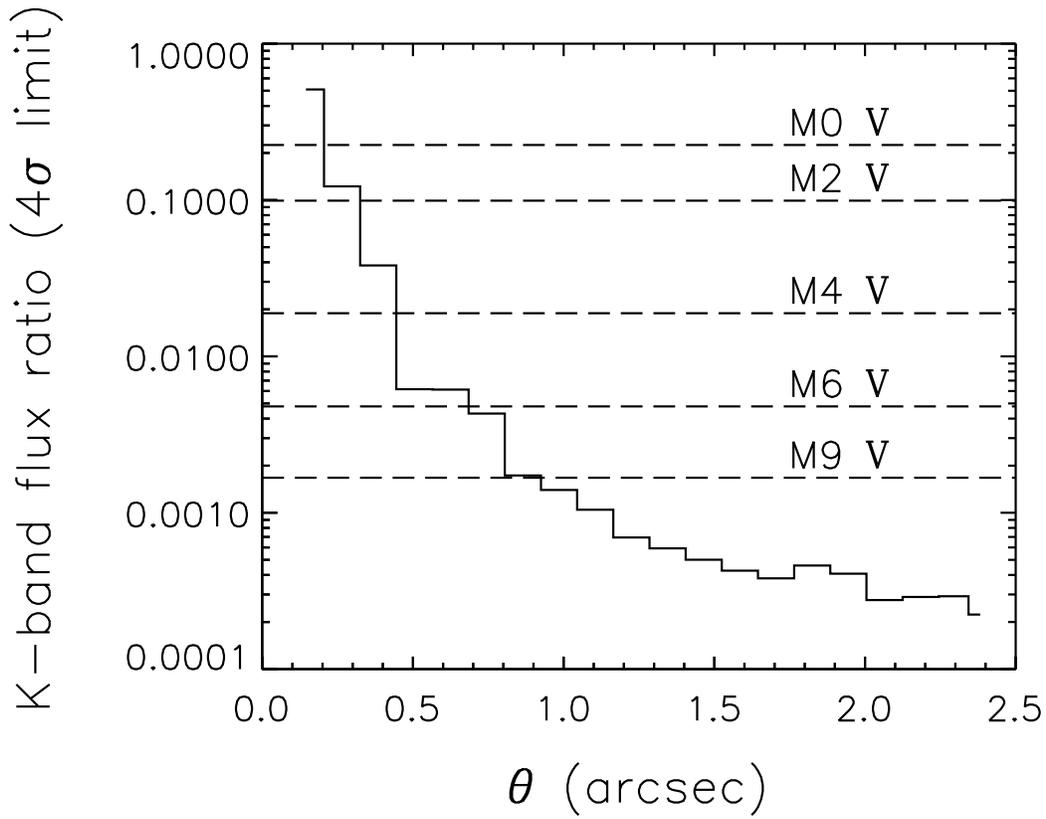,angle=90,width=6.5in}
\caption{Detectability of stellar companions near HD210277, based on
K--band (2.2 $\mu$) adaptive--optics images.  All main--sequence companions
between 17 and 250 AU (0.8--12 arcsec) would have been detected but
none was found.  The data rule out an M0V dwarf as close as 0.2'' (4.2 AU).}
\label{fig6}
\end{figure}

\begin{figure}
\psfig{figure=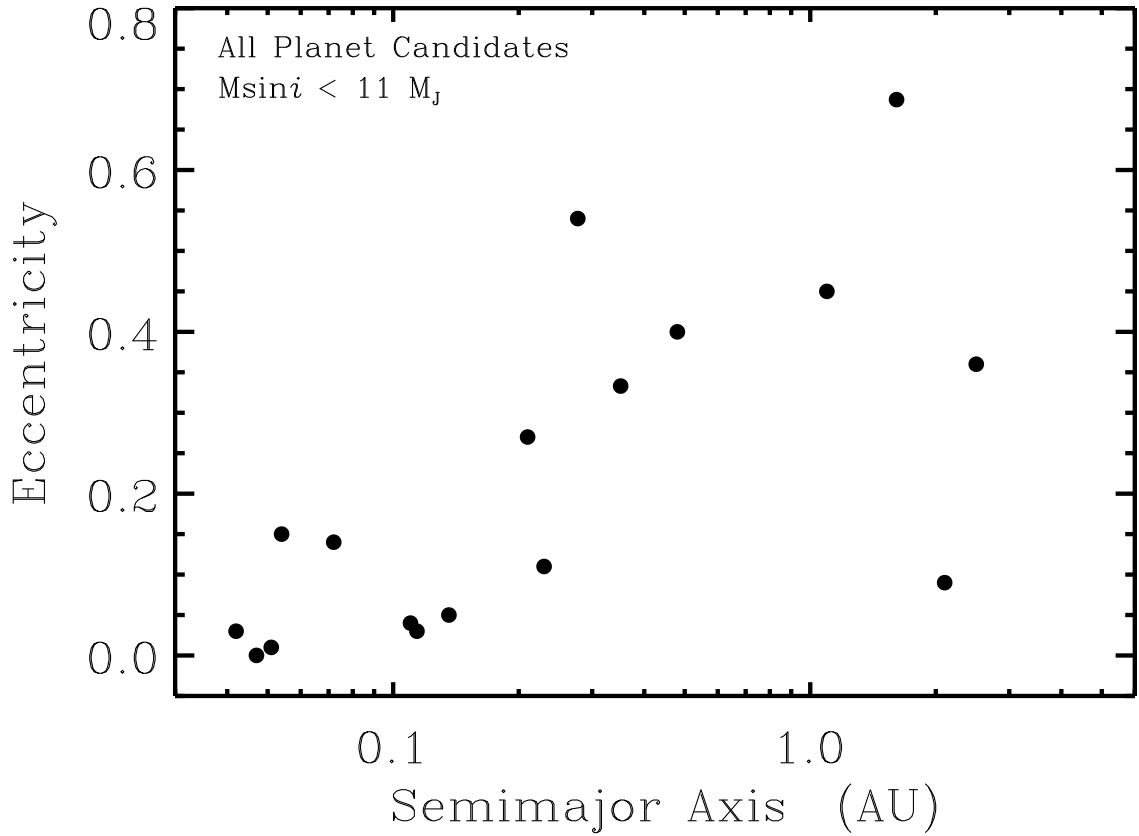,angle=90,width=6.5in}
\caption{Orbital eccentricity vs. semimajor axis for all 17 known
extrasolar planet candidates (\msini $<$ 11 \mjup).  
All small orbits are nearly circular, but all planet candidates
that have $a>$0.2 AU have eccentricities above that of Jupiter ($e$=0.05). }

\label{fig7}
\end{figure}

\begin{figure}
\psfig{figure=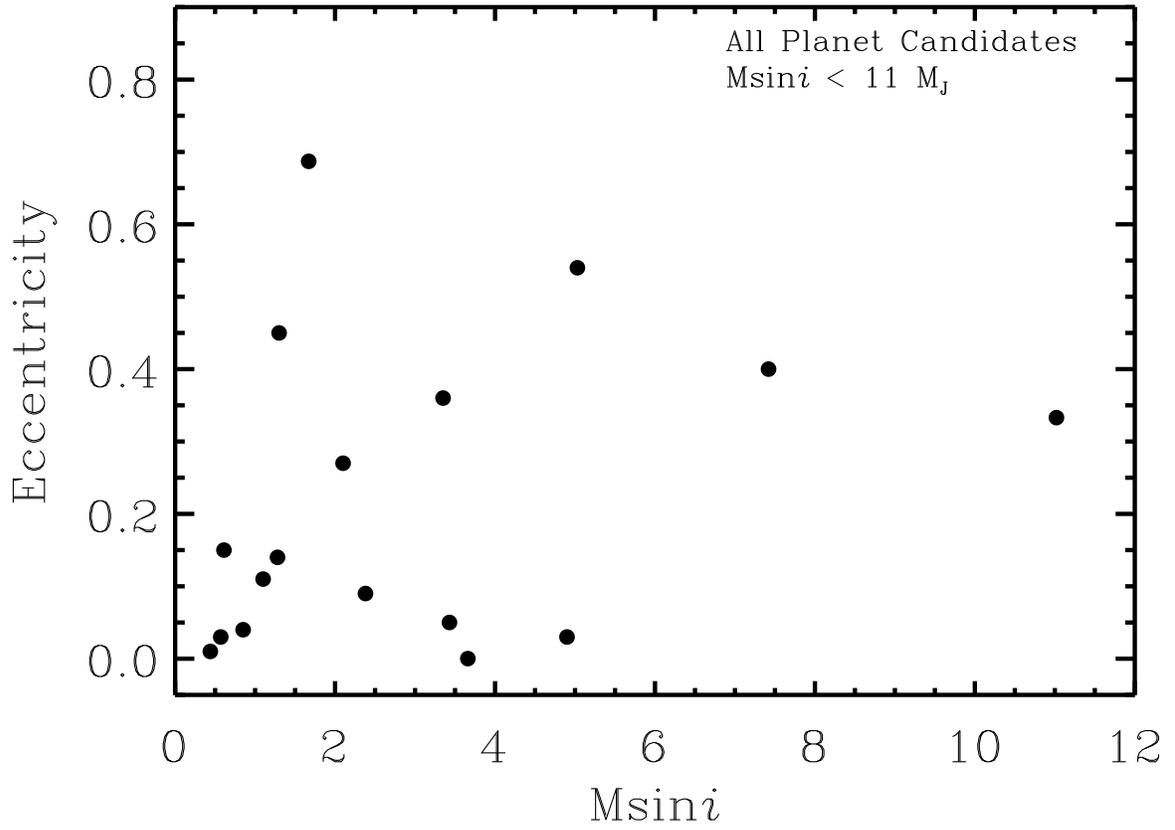,angle=90,width=6.5in}
\caption{Orbital eccentricity vs. \msini for all 17 known
extrasolar planet candidates that have \msini $<$ 11 \mjup.  
No trend with planet mass is apparent.}

\label{fig8}
\end{figure}

\begin{figure}
\psfig{figure=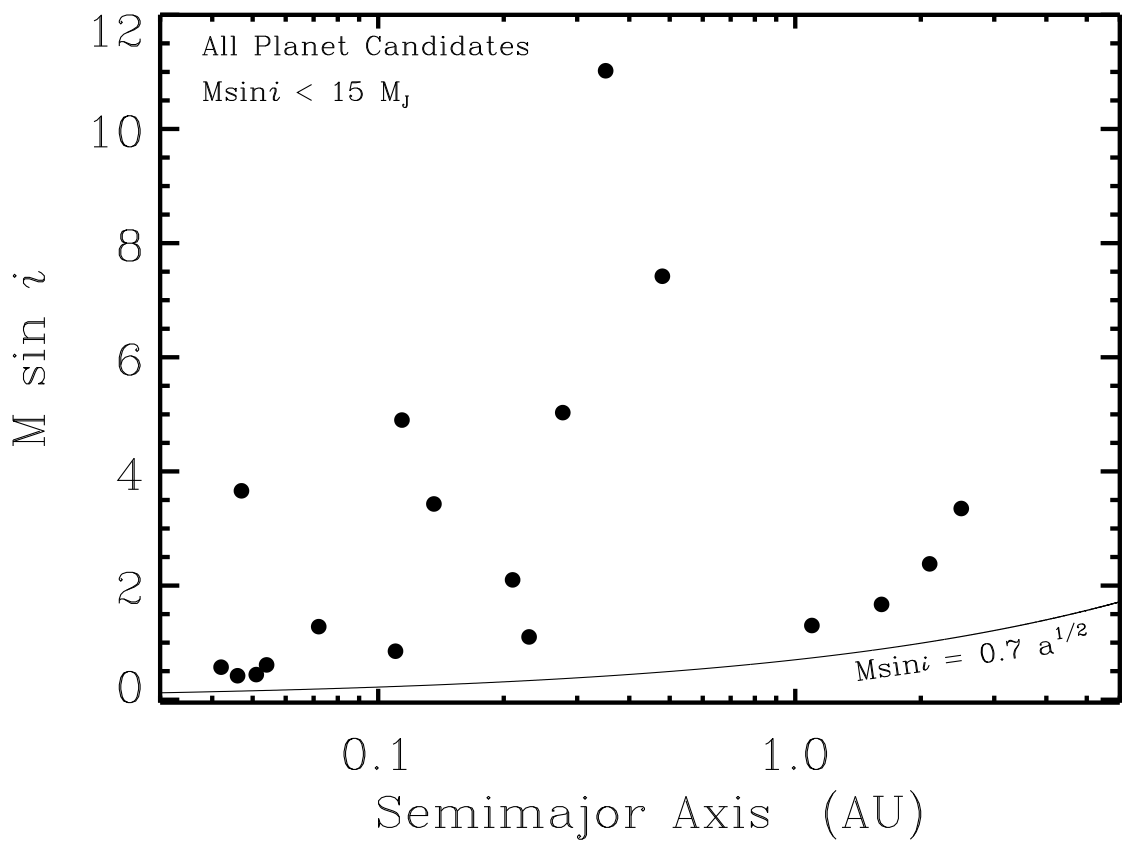,angle=90,width=6.5in}
\caption{\msini vs semimajor axis for all 17 extrasolar planet
candidates.  The lowest detectable values of \msini are shown as the
solid curve near the bottom (Cumming et al. 1999).  Planet candidates
are found at all values of semimajor axis from 0.05--2.5 AU.  The mass
distribution exhibits a cutoff at $\sim$6 \mjup, possibly the end of
the planetary mass function.  HD114762 appears above that prospective
mass limit.}
\label{fig9}
\end{figure}

\begin{figure}
\psfig{figure=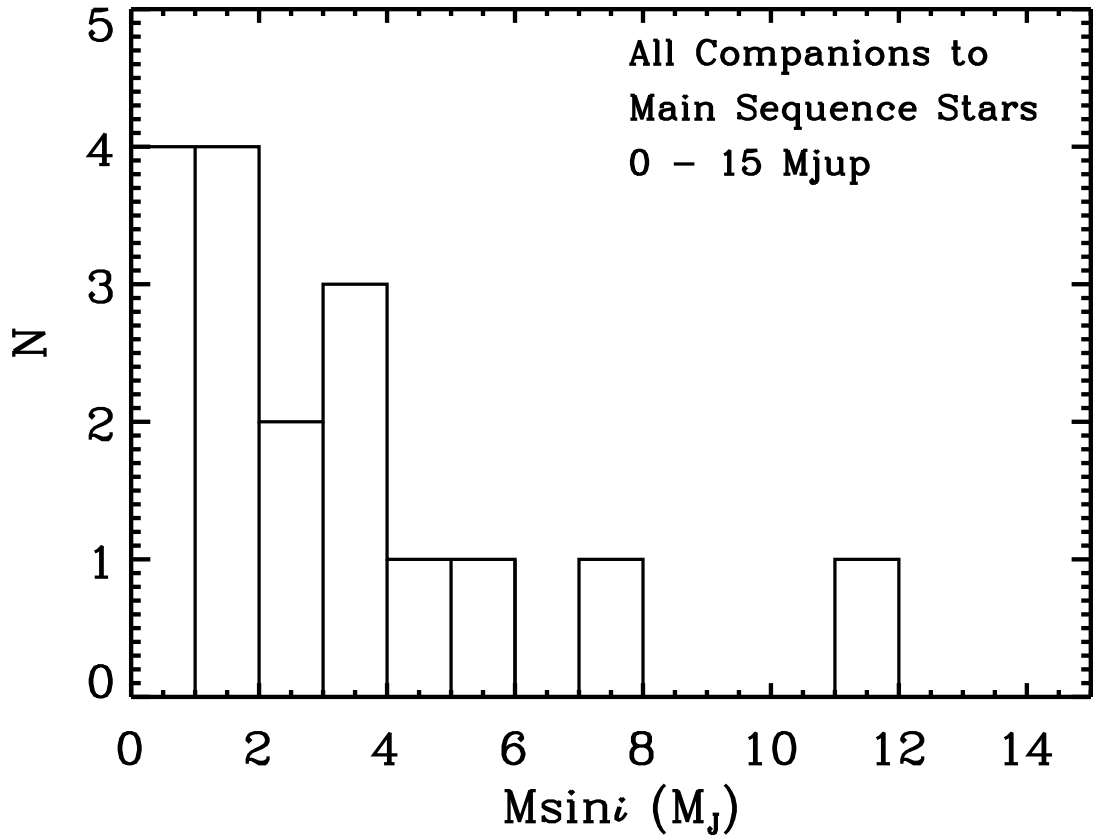,angle=90,width=6.5in}
\caption{Histogram of \msini in the range 0--15 \mjup 
for all known companions to main sequence stars.
The mass distribution exhibits steep drop for \msini$>$4 \mjup, indicating
a drop in the underlying companion mass function for $M > \sim$5 \mjup.}
\label{fig10}
\end{figure}

\clearpage

\begin{deluxetable}{crrrr}
\tablenum{1}
\tablecaption{Velocities for HD210277}
\label{vel210277}
\tablewidth{0pt}
\tablehead{
JD         &      RV         &  JD        & RV \\
-2450000   & $~~~$m s$^{-1}$ & -2450000   & $~~~$m s$^{-1}$ 
}
\startdata
 277.0404  &   15.8 &  983.0511  &   -8.4 \\
 366.7926  &   22.7 &  984.0878  &    0.0 \\
 418.7591  &   26.2 &  1010.0261  &  -40.2 \\ 
 462.7062  &   53.3 & 1011.1015  &  -36.4 \\
 605.0940  &  -23.3 & 1011.9692  &  -39.2 \\
 665.9876  &  -11.5 & 1013.0816  &  -39.0 \\
 688.9457  &    0.9 & 1014.0859  &  -39.6 \\
 689.9833  &    0.6 & 1043.0057  &  -35.4 \\
 713.8792  &    6.1 & 1043.9942  &  -32.0 \\
 714.9728  &   12.3 & 1050.9159  &  -24.7 \\
 715.9286  &   16.6 & 1051.9839  &  -35.0 \\
 783.7130  &   24.6 & 1068.8670  &  -14.0 \\
 784.7205  &   46.4 & 1069.9748  &  -18.0 \\
 785.6995  &   38.3 & 1070.9566  &  -20.2 \\
 805.7146  &   14.2 & 1071.8706  &  -16.6 \\
 806.7038  &   30.6 & 1072.9307  &  -17.2 \\
 956.0877  &   25.3 & 1074.8716  &   -3.8 \\
\enddata
\end{deluxetable}

\begin{deluxetable}{crrrr}
\tablenum{2}
\tablecaption{Velocities for HD168443}
\label{vel168443}
\tablewidth{0pt}
\tablehead{
JD & RV  & JD & RV \\
-2450000   &  $~~$m s$^{-1}$ & -2450000   &  $~~$m s$^{-1}$ 
}
\startdata
 276.9089  & -305.1 & 1010.8508  &   21.4 \\
 603.0118  &  -50.7 & 1011.8608  &   22.9 \\
 665.8678  &  -82.7 & 1012.9541  &   13.6 \\
 713.7377  &  -88.4 & 1013.0670  &   22.4 \\
 714.7665  &  -89.8 & 1013.8279  &    7.8 \\
 955.0104  &  -22.4 & 1013.9298  &    6.3 \\
 955.9586  &  -16.3 & 1042.9556  & -344.1 \\
 957.0711  &   -7.5 & 1043.9560  & -287.2 \\
 981.8801  & -578.4 & 1050.8141  &  -92.9 \\
 982.8913  & -505.7 & 1068.7704  &   39.1 \\
 983.0769  & -483.6 & 1069.7860  &   47.5 \\
 983.8223  & -412.4 & 1070.7981  &   40.2 \\
 984.0614  & -406.0 & 1071.7700  &   37.9 \\
1009.8701  &    8.1 & 1072.7627  &   33.3 \\
1010.0599  &   28.0 & 1074.7851  &   41.8 \\
\enddata
\end{deluxetable}

\begin{deluxetable}{crrr}
\tablenum{3}
\tablecaption{Orbital Parameters of HD210277 and HD168443}
\label{orbpar}
\tablewidth{0pt}
\tablehead{
Param                   & HD210277             & HD168443\tablenotemark{a} 
}
\startdata
P  (d)                  & 437 (25)            & 57.9 (1)          \\
${T}_{\rm p}$ (JD)    & 2450993 (20)      & 2450979.35 (2)        \\
e                       & 0.45 (0.08)           & 0.55 (0.04)     \\
$\omega$ (deg)        & 124 (20)              &  170 (5)          \\
K$_1$ (\ms)             & 41.5 (5.)          & 330. (23)         \\
a$_1 \sin i$ (AU)       & $1.49 \times 10^{-3}$  & $1.56\times 10^{-3} $ \\
f$_1$(m) (M$_\odot$)  & $2.29 \times 10^{-9}$ & $1.51 \times 10^{-7}$ \\
M$_2 \sin i$ (M$_{Jup}$) & 1.28 (0.4)           & 5.04 (0.4)         \\     
{\rm Nobs}          & 34                    & 30                     \\
\enddata
\tablenotetext{a}{Additional Velocity Slope is 89 $\pm$9 \ms per yr.}
\end{deluxetable}

\begin{deluxetable}{lrrrrr}
\tablenum{4}
\tablecaption{Orbital Parameters of Planet Candidates}
\label{candid}
\tablewidth{0pt}
\tablehead{
\colhead{Star}  & \colhead{M$_{\rm Star}$} & \colhead{a$~~$} & \colhead{P$~~$}   & \colhead{ecc.} & \colhead{M$\sin i$} 
\\
\colhead{ } & \colhead{(M$_{\odot}$)} & \colhead{\rm (AU)} & \colhead{\rm (days)} &\colhead{ }  & \colhead{(M$_{\rm J}$)}
} 
\startdata
HD187123 & 1.00 &  0.042 & 3.097    &0.03 & 0.57  \\
 Tau Boo & 1.20 &  0.047 & 3.3126   &0.00 & 3.66  \\
  51 Peg & 0.98 &  0.051 & 4.2308   &0.01 & 0.44  \\
 Ups And & 1.10 &  0.054 & 4.62     &0.15 & 0.61  \\
HD217107 & 0.96 &  0.072 & 7.11     &0.14 & 1.28  \\
  55 Cnc & 0.90 &  0.110 & 14.656   &0.04 & 0.85  \\
    GJ86 & 0.79 &  0.114 & 15.84    &0.04 & 4.90  \\
HD195019 & 0.98 &  0.136 & 18.3   &0.05 & 3.43  \\
   GJ876 & 0.32 &  0.210 & 60.9   &0.27 & 2.10  \\
 rho CrB & 1.00 &  0.230 & 39.6   &0.11 & 1.10  \\
HD168443 & 0.84 &  0.277 & 57.9   &0.54 & 5.04  \\
HD114762 & 0.82 &  0.351 & 84.0   & 0.334 & 11.02 \\
  70 Vir & 1.10 &  0.480 & 116.7  &0.40 & 7.42  \\
HD210277 & 0.92 &  1.097 & 437.   &0.45 & 1.28  \\
16 Cyg B & 1.00 &  1.61  & 803    &0.69 & 1.67  \\
  47 UMa & 1.03 &  2.09  & 1086   & 0.11 & 2.45  \\
  14 Her & 0.85 &  $>$2.50  & $>$2000 &0.36 & 3.35  \\
\enddata
\end{deluxetable}

\end{document}